\documentclass[preprint2]{aastex62}
\submitjournal{ApJ}

\shorttitle{Even simpler quadruple lens modeling}
\shortauthors{Schechter and Wynne}
\begin{document}

\title{Even simpler modeling of quadruply lensed quasars (and random
quartets) using Witt's hyperbola}

\correspondingauthor{Paul L.\ Schechter}
\email{schech@mit.edu}

\author[0000-0002-5665-4172]{Paul L.\ Schechter}
\affil{MIT Kavli Institute 37-664G,\\
  77 Massachusetts Avenue, Cambridge, MA, 02139-4307, USA}\
\author{Raymond A.\ Wynne}
\affil{MIT Department of Physics,\\
  77 Massachusetts Avenue, Cambridge, MA, 02139-4307, USA}\

%% Note that the \and command from previous versions of AASTeX is now
%% depreciated in this version as it is no longer necessary. AASTeX 
%% automatically takes care of all commas and "and"s between authors names.

%% AASTeX 6.2 has the new \collaboration and \nocollaboration commands to
%% provide the collaboration status of a group of authors. These commands 
%% can be used either before or after the list of corresponding authors. The
%% argument for \collaboration is the collaboration identifier. Authors are
%% encouraged to surround collaboration identifiers with ()s. The 
%% \nocollaboration command takes no argument and exists to indicate that
%% the nearby authors are not part of surrounding collaborations.

%% Mark off the abstract in the ``abstract'' environment. 
%\begin{abstract}
%\end{abstract}

%% Keywords should appear after the \end{abstract} command. 
%% See the online documentation for the full list of available subject
%% keywords and the rules for their use.
\keywords{{quasars --- gravitational lensing: strong}}

%% From the front matter, we move on to the body of the paper.
%% Sections are demarcated by \section and \subsection, respectively.
%% Observe the use of the LaTeX \label
%% command after the \subsection to give a symbolic KEY to the
%% subsection for cross-referencing in a \ref command.
%% You can use LaTeX's \ref and \label commands to keep track of
%% cross-references to sections, equations, tables, and figures.
%% That way, if you change the order of any elements, LaTeX will
%% automatically renumber them.
%%
%% We recommend that authors also use the natbib \citep
%% and \citet commands to identify citations.  The citations are
%% tied to the reference list via symbolic KEYs. The KEY corresponds
%% to the KEY in the \bibitem in the reference list below.

\begin{abstract}
    Witt (1996) has shown that for an elliptical potential, the
    four images of a quadruply lensed quasar lie on a rectangular
    hyperbola that passes through the unlensed quasar position
    and the center of the potential as well.  Wynne and Schechter (2018) have
    shown that, for the singular isothermal elliptical potential
    (SIEP), the four images also lie on an ``amplitude''
    ellipse centered on the quasar position with axes parallel to the
    hyperbola's asymptotes.  Witt's hyperbola arises from equating the
    directions of both sides of the lens equation.  The amplitude
    ellipse derives from equating the magnitudes.  One can model any
    four points as an SIEP in three steps.  1. Find the rectangular
    hyperbola that passes through the points.  2. Find the aligned ellipse
    that also passes through them. 3.  Find the hyperbola with
    asymptotes parallel to those of the first that passes through the
    center of the ellipse and the pair of images closest to each
    other.  The second hyperbola and the ellipse give an SIEP that
    predicts the positions of the two remaining images where the
    curves intersect.  Pinning the model to the closest pair
    guarantees a four image model.  Such models permit rapid
    discrimination between gravitationally lensed quasars and random
    quartets of stars.  
\end{abstract}

%\section{Four quasar images where a hyperbola intersects an ellipse}
\section{Introduction}

Wynne and Schechter (2018; henceforth WS) describe a robust scheme for
generating a singular isothermal elliptical potential (henceforth
SIEP) from the image positions of a quadruply lensed quasar.  It
converges even when the fit is poor or the model parameters
improbable.  It may be useful both in searching catalogs for
gravitationally lensed quasars (e.g. Delchambre et al 2019; Williams
et al 2017, 2018; Agnello et al 2018; and Lemon et al 2018) and in
providing a first guess for more sophisticated models (e.g.\ Keeton
2001).

Their method predicts images at the points of intersection of Witt's (1996)
hyperbola, and an ``amplitude'' ellipse.  The asymptotes of the
hyperbola, and the axes of both the potential and the amplitude
ellipse are all parallel to each other.  We call any frame with axes
parallel to these an ``aligned'' frame (as distinct from
the ``observed'' frame of the sky).

The WS scheme finds the amplitude ellipse iteratively, and has the
shortcoming that quartets of points sometimes result in SIEP models
that produce only two images.  We present here a simpler variant
of the WS approach that constructs an SIEP model without iteration and
which ensures four images.  It permits yet more rapid discrimination
between lensed quasars and random quartets of stars.

\section{The SIEP, the amplitude ellipse and Witt's hyperbola}

The two-dimensional effective potential $\psi$ (e.g. Schneider et al 1992)
for an SIEP centered on a lensing galaxy at $(x_g,y_g)$, is given by
\begin{equation}
\psi =  a^2 \left[{(x - x_g)^2 \over a^2} +{(y - y_g)^2 \over q^2 a^2}\right]^{1/2} \quad ,
\end{equation} 
in an aligned frame, where $q$ is the ratio of the
$y$ semiaxis to the $x$ semiaxis.  
The $x$ semiaxis, $a$, is either the the semi-major axis of the potential
if $q < 1$ or its semi-minor axis if $q > 1$.

The  amplitude ellipse is centered on the source at $(x_s,y_s)$
and is given by
\begin{equation}
{(x - x_s)^2 \over a^2} + {(y - y_s)^2 \over a^2/q^2} = 1
\end {equation}
in an aligned frame, where $1/q$ is the ratio of its y-semiaxis to its
x-semiaxis.  It is orthogonal to the elliptical potential and has the
same shape, but has a larger semi-major axis.

The hyperbola is offset from both the potential and the amplitude
ellipse.  For an elliptical potential in an aligned frame
it is given by Witt's equation,
\begin{equation}
(x - x_g)(y - y_s) = {1 \over q^2}(y -  y_g)(x - x_s) \quad ,
\end{equation}
from which one sees that both the lensing galaxy and the source lie on
the hyperbola.  The coordinates of the center of the hyperbola,
$(x_h,y_h)$, are found to be
\begin{equation}
x_h = {(-y_s + q^2y_g) \over (1 - q^2)}  \quad ; \quad
    y_h = {(-x_s + q^2x_g) \over (1 - q^2)}      
\end{equation}
in an aligned frame.  The semi-major axis of the hyperbola,
is equal to $\sqrt{|c^2|}$, where
\begin{equation}
c^2 \equiv -2{q^2\over 1-q^2}(x_s - x_g)(y_s - y_g) \quad .
\end{equation}
The hyperbola's semi-major axis is parallel to the line $y=x$ if $c^2
> 0 $ and is otherwise parallel to the line $y = -x$.  The hyperbola
collapses to two perpendicular lines when the displacement of the
source from the center of the potential is parallel to either of the
axes of an aligned frame.  Witt's equation holds for {\it all}
elliptical potentials.

In what follows we describe a variant of the WS scheme
that is even simpler by virtue of solving directly for the amplitude ellipse
rather than iteratively.  It also reproduces, by construction, the
separation of the closest pair of images, upon which their predicted
fluxes strongly depend.

\section{The method}

We first present our recipe, then elaborate on it and finally explain it.

\subsection{The recipe}

\begin{enumerate}
\item Find the rectangular hyperbola  passing through the four image
    positions in the observed frame;

\item find the image coordinates in an ``aligned'' frame
    with axes parallel to the asymptotes of the hyperbola;

\item find the amplitude ellipse passing through
    the four image positions and aligned with these axes;

\item find the two images subtending the smallest angle from
    the center of the amplitude ellipse;

\item evaluate Witt's equation at  the positions of these two images 
    to solve for the unknown  coordinates of the lensing galaxy, $(x_g,y_g)$.
\end{enumerate}    

\begin{figure*}[t]
  \plotone{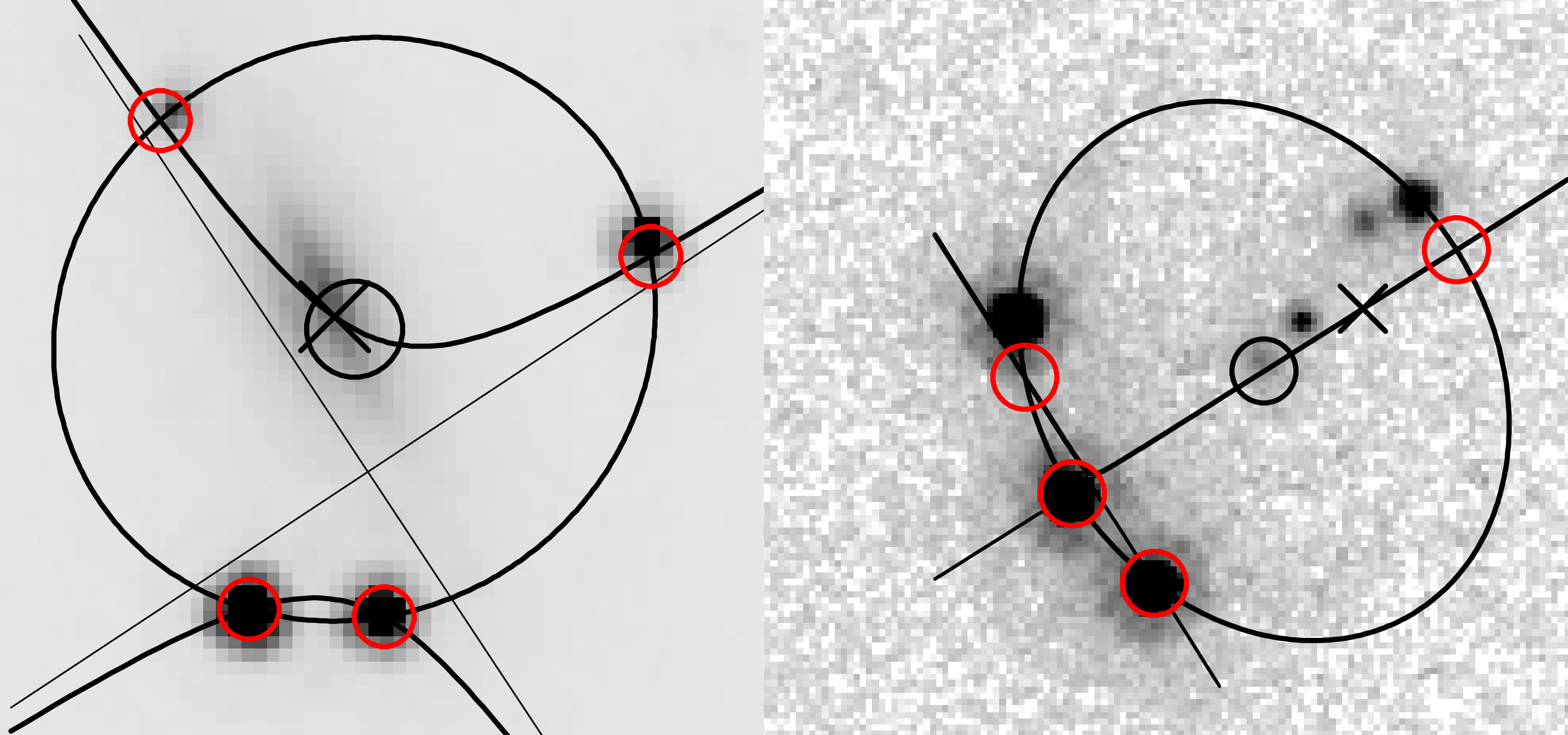}
  \caption{HST F814W images of SDSSJ1330+1810 and PSJ0630-1210.  The
    ellipses and hyperbolae for the models are shown, with images
    predicted where they intersect.  A black circle marks the source
    position and an ``x'' marks the galaxy position.  By construction
    the model exactly reproduces the closest image pair.  The
    semi-major axes of the ellipses are $0\farcs97$ and $1\farcs44$
    respectively. The observed position of the galaxy in
    SDSSJ1330+1810 lies close to the inferred position, and its
    orientation is likewise close to the model's.  PSJ0630-1210 has
    two lensing galaxies, one very close to the inferred source position and
    one just inside the ellipse near the northernmost image.  A fifth
    quasar image lies between the two galaxies. The hyperbola has so
    small a semi-major axis, $0\farcs04$, that it merges with its
    asymptotes.}
\end{figure*}

\subsection{Elaboration}

Note that Witt's equation {\it again} describes a hyperbola, with the
source and the galaxy on its ``primary'' branch.  By construction the
two closest images lie on the secondary branch.  The other two images
are offset from the primary branch.  We take the
rms offset of the four images from their predicted positions
(normalized by the semi-major axis of the amplitude ellipse), as a
figure of merit, $G$, for the SIEP model.

The WS SIEP model differs from the present one in that their hyperbola
passes through all four images, and their iteratively fit ellipse
passes through none (except when the fit is perfect).  In the present,
non-iterative scheme, it is the {\it ellipse} that passes through all
four images, with our {\it final} hyperbola passing through only two.
It is faster by virtue of its non-iterative nature.  Moreover, it
reproduces, by construction, the separation of the closest pair of
images, which has a strong influence on their predicted fluxes.

The rectangular hyperbola and the amplitude ellipse are found using
their representations as conic sections, $Ax^2 + Bxy + Cy^2 + Dx + Ey
+ F = 0$, in the observed and aligned frames, respectively.  The
rectangular hyperbola of the first step has coefficients $A_h$ and
$C_h$ equal and opposite. The coefficient $B_h$ may be set equal to
unity.  Evaluating the conic at the four image positions gives
equations that are linear in the four remaining unknowns, $A_h$,
$D_h$, $E_h$ and $F_h$.  In the present scheme only $A_h$ is used,
giving the angle that an aligned frame makes with the observed frame,
\begin{equation}
\theta = -{1 \over 2} \arctan{2 A_h}
\end{equation}

The amplitude ellipse will have $A_e = 1$ and $B_e = 0$ in an aligned
frame, again giving four equations linear in four unknowns, $C_e$,
$D_e$, $E_e$, and $F_e$ when evaluated at the four image positions.
If $C_e < 0$, the resulting conic section is a hyperbola, not an
ellipse, and is unlikely to have resulted from an SIEP-like potential.
This was the case for roughly one third of the random configurations
considered below.  We extract the square of the axis ratio of the
potential, $q^2 = C_e$.  

Following Witt's example, one can further simplify the calculation by
taking the origin of the ``observed'' frame to be one of the four
images, in which case the coefficient $F_h = 0$.  One finds the
hyperbola by evaluating the conic at the other three images.  This
trick likewise simplifies solving for the amplitude ellipse if one
rotates about Witt's origin to transform from the observed
frame to the the aligned frame.  The coefficient $F_e$ is then zero
and one solves for the amplitude ellipse by evaluating the conic at the other
three images.

\subsection{Explanation}

Both the hyperbola and the amplitude ellipse derive from the ``lens equation''
(Schneider et al 1992), which sets the image deflection equal to the
gradient of a two-dimensional gravitational potential.  Witt's
hyperbola is obtained by equating the directions of the two sides of
the lens equation.  The amplitude ellipse is obtained by equating the magnitudes
of the two sides.  A proper solution of the lens equation must satisfy
both components.  {\it Any} elliptical potential will have its images,
source and galaxy on a rectangular hyperbola (Witt 1996), but only for the
SIEP do the images also lie on an associated ellipse with the source
at the center.

\begin{figure}
  \plotone{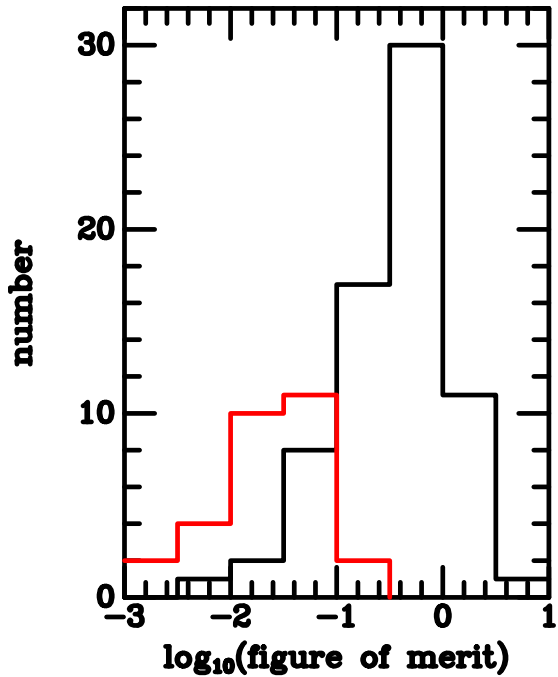}
  \caption{The red line shows
    the distribution of the figure of merit $G$ for the 29 known lenses in
    WS. The black line shows the same distribution for 70 random quartets.
    The remaining 30 could not be modeled as an SIEP.}
\end{figure}

\section{Application}

\begin{figure*}
  \plotone{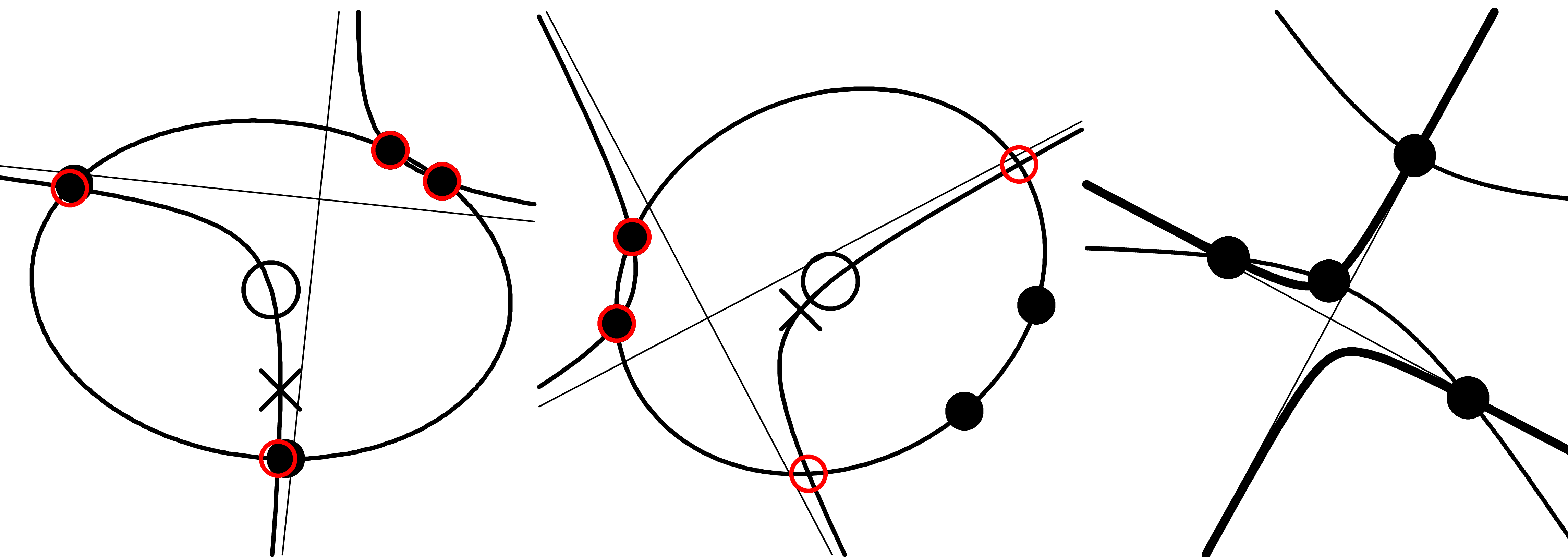}
  \caption{Three random quartets of points.  The first quartet
    is fit unusually well by the SIEP model.  The second quartet
    is typical of those for which our procedure yielded an SIEP
    model.  The third is typical of those for which the conic
    section of our third step gave a hyperbola (thin lines)
    rather than the anticipated amplitude ellipse.  The rectangular
    hyperbola from our first step (thicker lines) has three
    images on one branch and one on the other, rather than two
    on each.}
\end{figure*}

\subsection{Known quadruply lensed quasars}

We have applied the above technique to the 29 systems analyzed in WS.
These yielded SIEP models very similar to those in WS for the better
fitting systems but agreed less well for those with poorer fits.

In Figure 1 we show the result of applying the present technique to
the gravitationally lensed quasars SDSSJ1330+1801 (Oguri et al 2008)
and PSJ0630-1201 (Ostrovski et al 2017), with positions taken from
Shajib et al (2019).  These gave two of the three worst WS fits, with
their amplitude ellipses intersecting only the primary branch of the
hyperbola, producing just two images.  Random
quartets exhibit this pathology more frequently.  Source plane fitting
algorithms, which minimize the scatter in the positions inferred for
the source for each of the four images, can also predict two rather
than four images.

This renders the WS approach ill-suited to using observed fluxes to
further discriminate between quadruply lensed quasars and random
quartets.\footnote{This shortcoming would be
  remedied by constraining the WS fit of the amplitude ellipse to pass
  through the two closest images.}  By construction the present method
produces four images, with the distance between the two closest
images exactly as observed.  Close
pairs indicate a source that lies near a fold caustic (Gaudi and
Petters 2002), with the magnification of the pair varying inversely as
the distance between them.

\subsection{Random quartets}

We attempted to produce SIEP models for the same 100 random quartets
analyzed in WS.  As noted above, no amplitude ellipse could be drawn
through roughly 1/3 of these, with the third step yielding a hyperbola
instead.  In Figure 2 we show a histogram of the figure of merit $G$
for the surviving random quartets and for the 29 known lenses.  While
the present figure of merit is roughly a factor of two smaller than
that of WS, for both the known lenses and the random quartets,
they produce similar rank orderings.  If we reject systems with
$G > 0.1$ we lose two known lenses (both of which have two lensing
galaxies) and accept eleven random quartets.  If we further reject
systems with axis ratios outside the interval $0.4 < q < 2.5$, the
false positive rate drops to 8\%  without losing any more known lenses,
  the most extreme of which, 2M1134-2103, has $q = 0.49$.

Lucey et al (2017) attribute the quadrupole moment in 2M1134-2103 to
an external shear, $\gamma = 0.34$, rather than to an elliptical {\it
  mass} distribution, which would have an axis ratio $q_m \sim 0.1$.
Using Keeton's (2001) {\tt lensmodel} program we have generated many
synthetic quartets assuming a singular isothermal sphere with external
shear.  We consistently find $q = (1-\gamma)/(1 + \gamma)$, with the
SIEP giving {\it perfect} fits to these synthetic quartets.  Our
cutoff of $q = 0.4$ corresponds to a shear $\gamma = 0.429$, larger
than any known for a lensed quasar.

In Figure 3 we show models for three random quartets of points.  The
first quartet has $G = 0.028$, better than most of the known lenses.
The second has $G = 0.598$, typical of the random quartets and
substantially worse than the known lenses.  The third quartet was
among those for which no SIEP model could be found. For this system we
show the hyperbola derived from the first step and the unwanted
hyperbola derived from the attempt to find an amplitude ellipse.

For an SIEP, the amplitude ellipse is centered on the primary branch
of Witt's hyperbola, and must intersect it in at least two points.  It
may or may not intersect the secondary branch, but will do so twice when it
does.  Yet for the third quartet, three of the four random points lie on
one branch of Witt's hyperbola.  It is no surprise that an SIEP model
could not be found

\section{Using fluxes to improve discrimination between quadruply
  lensed quasars and random quartets}

Delchambre et al (2019) catalog 70,000 quartets of GAIA point sources,
from which they recover eleven known quadruply lensed quasars and find
at least one new confirmed quad.\footnote{The probability of finding a
  random quartet increases dramatically at low galactic latitude, with
  more than 98\% of their catalogued quartets at $|b| < 30^\circ$.}
Their ``extremely randomized trees'' method (ERT) does better at
finding these than the present approach, with the twelve confirmed
lenses ranked higher than 1440 and 5600 in their ranked list of quads
and ours, respectively.  This may be explained by the implicit use of
both flux ratios {\it and} positions by the ERT algorithm, where the
present method uses only positions.

While the second and third random quartets in Figure 3 are not likely
to be mistaken for a lensed quasar, the first quartet is an excellent
impostor if one considers only positions.  But if one uses the SIEP
model to predict fluxes, one finds that the two close images are
highly magnified and very nearly equal in magnitude.  The next
brightest is 1.5 magnitudes fainter, and the least bright, close to
the lensing galaxy, is 4 magnitudes fainter.  While the flux ratios
for a random quartet of stars will depend upon its Galactic
coordinates and the depth of the survey, they are unlikely to match
this pattern.

Unfortunately, fluxes for the individual macro-images predicted by an
SIEP model are subject to micro-lensing by the stars in the lensing
galaxy (Paczy\'nski 1986).  Therefore SIEP-predicted fluxes may not
improve discrimination as much as might otherwise be thought.
Macro-images may deviate by two magnitudes or more depending upon the
convergences and shears at their positions (which are known
from the SIEP model) and the surface mass density of micro-lenses
(Wambsganss 1992), which even under the simplest of assumptions
depends upon the redshifts of the source and lens.  Taking these
effects into account is more computationally intensive than the
present discrimination based only on positions.

A further complication in using fluxes is our inability to distinguish
between an SIEP configuration and an externally sheared singular
isothermal sphere when considering only positions.  The two
alternatives give flux ratios that can vary by 20-30\%.  Moreover, the
absolute magnifications are to first order a factor of two greater in
the case of external shear, so the effects of micro-lensing will be
different in the two cases.

These complications notwithstanding, flux ratios are likely to
provide at least some improvement in discriminating between
quadruply lensed quasars and random quartets.

%% The "ht!" tells LaTeX to put the figure "here" first, at the "top" next
%% and to override the normal way of calculating a float position
%\begin{figure}[ht!]

%% If you wish to include an acknowledgments section in your paper,
%% separate it off from the body of the text using the \acknowledgments
%% command.
\acknowledgments
We are indebted to Chuck Keeton and Alar
Toomre for crucial suggestions, to an anonymous referee for
helpful comments, to Anowar Shajib for the
HST images and to Armin Deutsch (1966) for inspiration.
\vfill
\eject

\end{document}